\documentclass[prb,
twocolumn,
superscriptaddress,showpacs,amsmath,amssymb]{revtex4}

\begin{document}

\title{Spin vortex lattice in the Landau vortex-free state of rotating superfluids}

\author{G.E.~Volovik}
\affiliation{Low Temperature Laboratory, Aalto University,  P.O. Box 15100, FI-00076 Aalto, Finland}
\affiliation{Landau Institute for Theoretical Physics, acad. Semyonov av., 1a, 142432,
Chernogolovka, Russia}

\date{\today}

\begin{abstract}
We show that the Landau vortex-free state in rotating container may give rise to the lattice of spin vortices. We consider this effect on example of spin vortices in magnon Bose-Einstein condensate (the phase coherent spin precession) in the B-phase of superfluid $^3$He, and on example of spin vortices in the polar phase of $^3$He.
\end{abstract}
\pacs{
}

\maketitle

\section{Introduction}

In the rotating vessel the lattice of mass vortices represents the ground state or the thermal equilibrium state of the rotating superfluid.
For spin vortices the situation is different. Orbital rotations does not act on the spin vortices, and the other external or internal fields are needed for the formation of the lattice, such as Dzyaloshinskii-Moriya interaction,\cite{Dzyaloshinskii1958,Moriya1960} which leads to the formation of skyrmion lattice, or some deformations of the crystal.
Here we show that the lattice of spin vortices can be created in rotating vessel, if the formation of the mass vortices is suppressed,.
The Landau vortex-free state in the rotating vessel (the analog of Meissner state in superconductors) acts on spin vortices in the same way as rotation acts on mass vortices.  

\section{Mixture of spin and mass superfluids}

The hydrodynamics of mixture of two superfluids contains the term in the hydrodynamic energy, which mixes two superfluid velocities
\cite{AndreevBashkin1975,VolovikMineevKhalat1975}. Here we consider magnon BEC in superfluid  $^3$He-B -- the homogeneously precessing domain (HPD)\cite{HPDexp,Fomin1984,Fomin1985}. When  the spontaneous coherent spin precession takes place in the moving superfluid  $^3$He-B, we  have the mixture of spin superfluidity of magnon BEC and mass superfluidity of  $^3$He-B. 

The magnon BEC or the phase coherent precession of magnetization is characterized by the density of magnons in the condensate  $n_{\rm M}=S(1-\cos\beta)=S-S_z$, where $S=\chi H$ is spin density in magnetic field; $S_z$ is the projection of the precessing spin on magnetic field; and $\beta$ is the tilting angle of the precessing magnetization, see review \cite{MagnonBEC}. The magnon BEC has the superfluid velocity
\begin{equation}
{\bf v}_{\rm M}=\frac{\hbar}{m_{\rm M}}\boldsymbol{\nabla}\alpha \,,
\label{MagnonSuperfluidV}
\end{equation}
where $\alpha$ is the angle of the coherent precession, which plays the role of the phase of magnon BEC; and $m_{\rm M}$ is magnon mass (for simplicity we neglected the anisotropy of this mass).
In the spin dynamics of magnon BEC, the magnon density $n_{\rm M}$ and the phase $\alpha$ are canonically conjugate variables, and thus 
\begin{equation}
{\bf P}=n_{\rm M}\boldsymbol{\nabla}\alpha =m_{\rm M}n_{\rm M}{\bf v}_{\rm M}\,,
\label{MagnonMomentum}
\end{equation}
represents the momentum density of the magnon field.
 
Since magnon BEC is the excited state of the background superfluid, in the moving superfluid the magnon BEC acquires the Doppler shift energy term:
\begin{equation}
F_{\rm mix}={\bf P}\cdot({\bf v}_{\rm s}-{\bf v}_{\rm n})= m_{\rm M}n_{\rm M}{\bf v}_{\rm M}\cdot({\bf v}_{\rm s}-{\bf v}_{\rm n}) \,.
\label{MixTerm}
\end{equation}
It describes the interaction of magnon BEC with the mass flow in the $^3$He-B, where ${\bf v}_{\rm s}$ and ${\bf v}_{\rm n}$ are correspondingly superfluid and normal velocities of the B-phase.  
This term, which mixes superfluid velocity of the background mass superfluid and superfluid velocity of magnon BEC, represents another realization of the Andreev-Bashkin effect in superfluid mixtures, when the superfluid current of one component depends on the superfluid velocity of another component.\cite{AndreevBashkin1975,VolovikMineevKhalat1975}
On the other hand the counterflow ${\bf v}_{\rm s}$ and ${\bf v}_{\rm n}$ together with spin density plays the similar role as  Dzyaloshinskii-Moriya interaction in magnets, which violates the space inversion symmetry. In magnets this leads to formation of skyrmion lattices, see e.g. \cite{Kurumaji2019}.

The mixed term modifies the kinetic energy of magnon BEC:
\begin{equation}
F_{\rm grad}= \frac{1}{2}m_{\rm M}n_{\rm M}{\bf v}_{\rm M}^2   + m_{\rm M}n_{\rm M}{\bf v}_{\rm M}\cdot({\bf v}_{\rm s}-{\bf v}_{\rm n}) \,.
\label{HydrodynamicEnergy}
\end{equation}
This equation is valid in the zero temperature limit, where however the  HPD experiences the Suhl instability.\cite{Bunkov2006} 
For finite temperatures, at which the HPD is stable, the mixed term is smaller, but it has the same order of magnitude.

\section{Formation of spin vortex lattice in magnon BEC}

Now let us consider the Landau state in the container rotating with angular velocity $\boldsymbol{\Omega}$. This is the state in which the normal component of the liquid has the solid body rotation with velocity ${\bf v}_{\rm n}= \boldsymbol{\Omega} \times {\bf r}$, while the superfluid component of the B-phase is vortex-free and thus is not rotating, ${\bf v}_{\rm s}=0$. In $^3$He-B the Landau state  is typical, because the critical velocity of the creation of quantized vortices is large due to the large energy barrier, which is necessary to overcome for the creation of mass vortices.\cite{Parts1995}
In this case the gradient energy of magnon BEC (omitting the constant term) becomes
\begin{equation}
F_{\rm grad}= \frac{1}{2}m_{\rm M}n_{\rm M} \left({\bf v}_{\rm M}- \boldsymbol{\Omega} \times {\bf r}\right)^2 \,.
\label{HydrodynamicEnergy2}
\end{equation}
This means that the Landau state in the rotating vessel acts on spin superfluid (magnon BEC) in the same way as rotation acts on mass superfluid, i.e. it should lead to formation of spin vortices, in which the phase $\alpha$ has $2\pi$ winding (single spin vortex in magnon BEC has been observed in Ref.\cite{SpinVortex1990}).
So, if the creation of mass vortices is suppressed, but the creation of spin vortices is allowed, one obtains the state with the lattice of spin vortices. 

The number of these spin vortices in the Landau state in rotating vessel is determined by the circulation quantum of spin vortex 
$\kappa_{\rm M}=2\pi \hbar/m_{\rm M}$ and by angular velocity. That is why the number of spin vortices in the lattice in the Landau state can be expressed in terms of the equilibrium number of quantized mass vortices in the fully equilibrium rotating state:
\begin{equation}
\frac{N_{\rm spin}}{N_{\rm mass}} =\frac{\kappa_3}{\kappa_{\rm M}}
=\frac {m_{\rm M}}{2m_3}  \,,
\label{NumberVortices}
\end{equation}
where $\kappa_3= 2\pi \hbar/2m_3$ is the quantum of circulation in superfluid $^3$He-B and $m_3$ is the mass of $^3$He atom. 
In this equation we compared two rotating states: the fully equilibrium rotating state where the mass vortices form the vortex lattice, while the spin vortices are absent; and the metastable Landau state in rotating vessel, where mass vortices are absent, while spin vortices form the lattice.

In the isotropic approximation, which we used,  the magnon mass is $m_{\rm M}=\omega_L/2c_{\rm s}^2$, where $\omega_L$ is Larmor frequency, and $c_{\rm s}$ is the speed of spin waves in $^3$He-B.\cite{MagnonBEC}
In typical experimental situations one has $N_{\rm M} <1$, and thus spin vortices were not created in the Landau state.
The spin vortex lattice could become possible in the higher magnetic fields.
The condition for $N_{\rm M} >1$ should be consistent with condition that the mass vortices are still not created and thus the mass counterflow persists. For that the Feynman critical velocity for formation of spin vortex should be smaller than the critical velocity $v_c$ at which the energy barrier for formation of the mass vortex vanishes: 
\begin{equation}
\frac{\hbar}{m_{\rm M}R} <v_c \,.
\label{Condition}
\end{equation}
Here $R$ is the radius of the container. 

\section{From helical spin texture to spin vortex lattice}

Similar effect of the formation of the vortex lattice takes place can be applied  to the conventional spin vortices in superfluid $^3$He. The main problem there is that the spin vortices are influenced by the spin-orbit interaction, due to which they become the termination lines of the topological solitons.\cite{MineyevVolovik1978}
Spin vortices with soliton tail have been observed in $^3$He-B.\cite{Kondo1992} 

In the magnon BEC discussed in the previous sections the spin-orbit problem is absent: in the precessing state the spin-orbit interaction is averaged over the fast precession, and as a result the spin vortices have no solitonic tails, and thus may form the lattice in the Landau state. The solitonic tails appear in the applied radio-frequency field, where the energy of the condensate explicitly depends on $\alpha$. But in the free precession the spin vortices are free.

For the conventional spin vortices, the spin-orbit problem is resolved in the polar phase of $^3$He, which exists in the nano-scale confinement (in the so called 
nafen).\cite{Aoyama2006,Askhadullin2012,Dmitriev2015,Dmitriev2018,Dmitriev2019,Halperin2018,Volovik2018,Nissinen2017,Ikeda2020} If the magnetic field ${\bf H}$ is parallel to the strands of nafen, the solitonic tails are absent. As a result both the spin vortices and Alice strings (the objects, which combine the half-quantum mass vortex and the half-quantum spin vortex) 
have been observed.\cite{Autti2016,Makinen2019,Kuang2020} The Landau states have been also observed in the polar phase in spite of zero value of the Landau critical velocity in this superfluid with Dirac nodal line.\cite{Autti2020}
So, let us consider  the Landau state in the polar phase.

According to Brauner and Moroz,\cite{Brauner-Moroz2019} in the presence of both the counterflow ${\bf v}_{\rm s}-{\bf v}_{\rm n}$ and magnetic field ${\bf H}$, the spin texture is formed, which they call the helical spin texture. The spin texture originates form the similar mixed term in Eq.(\ref{MixTerm}):
\begin{equation}
F_{\rm mix}=S \boldsymbol{\nabla}\alpha\cdot({\bf v}_{\rm s}-{\bf v}_{\rm n}) \,,
\label{Mixed}
\end{equation}
where $\alpha$ is the angle of the unit $\hat{\bf d}$-vector,  which describes the spin part of the order parameter, and  $S=\chi H$ is spin density in magnetic field. Let us mention that Brauner and Moroz considered the formation of the  $\hat{\bf d}$-texture in a different phase -- in the chiral A-phase. Both the A-phase and the polar phase belong to the class of spin-triplet $p$-wave superfluids with equal spin pairing.\cite{VollhardtWolfle1990} In both phases the spin part of the order parameter is described by the $\hat{\bf d}$-vector, but in the A-phase the spin-orbit problem exists. To avoid the effect of spin-orbit interaction, Brauner and Moroz considered the quasi-two-dimensional system -- thin film of $^3$He-A. 

Extension of the constant counterflow discussed by  Brauner and Moroz to the Landau state of the polar phase in the rotating cryostat, 
with ${\bf v}_{\rm n}= \boldsymbol{\Omega} \times {\bf r}$ and ${\bf v}_{\rm s}=0$, is straightforward. The gradient energy for spin textures becomes (omitting the constant term):
\begin{equation}
F_{\rm grad}= \frac{1}{2}\rho_{\rm spin} \left(\boldsymbol{\nabla}\alpha- \frac{S}{\rho_{\rm spin} }\boldsymbol{\Omega} \times {\bf r}\right)^2 \,.
\label{HydrodynamicEnergy3}
\end{equation}
This corresponds to Eq.(\ref{HydrodynamicEnergy2}) with magnon mass $m_{\rm M}= S/\rho_{\rm spin}$, and thus instead of the helical texture suggested in Ref.\cite{Brauner-Moroz2019} one obtains the lattice of spin vortices. The number of equilibrium spin vortices in the Landau state in the vessel of radius $R$ is:
\begin{equation}
N_{\rm spin} =\frac{\chi H}{\rho_{\rm spin}}\Omega R^2 \,.
\label{NumberVortices}
\end{equation}
It has the same order of magnitude as in the case of magnon BEC, and also requires large magnetic field for the experimental realization of spin-vortex lattice.

\section{Conclusion}

For the spin-triplet superfluids, such as superfluid $^3$He,\cite{VollhardtWolfle1990,Mizushima2016} the Landau state of the superfluid in the rotating container can be the source of the formation of the vortex lattice of spin vortices. In the polar phase of $^3$He, the lattice of the conventional spin vortices is possible. In the B-phase of superfluid $^3$He, the lattice of vortices can be formed in the magnon Bose condensate. The spin-vortex lattice is formed, if the Feynman critical velocity for the creation of a spin vortex is lower than the real critical velocity for creation of the mass vortex. In superfluid $^3$He the latter critical velocity is rather large, because of the large energy barrier. 

Similar phenomenon may occur in rotating neutron stars (review on superfluidity and superconductivity in neutron stars see in Ref.\cite{Sedrakian2018}).

  {\bf Acknowledgements}. This work has been supported by the European Research Council (ERC) under the European Union's Horizon 2020 research and innovation programme (Grant Agreement No. 694248). I thank Vladimir Eltsov for discussions.

\end{document}